\documentclass[aps,pra,floats,floatfix,superscriptaddress,twocolumn]{revtex4-2}
\usepackage{ifthen}
\usepackage{ifpdf}
\usepackage{color}
\ifpdf
\usepackage{graphicx}
\usepackage{epstopdf}
\else
\usepackage{graphicx}
\usepackage{epsfig}
\fi
\usepackage{ulem}
\usepackage{latexsym}
\usepackage{amsmath}
\usepackage{amssymb}
\usepackage{bm}
\usepackage{wasysym}

\usepackage{latexsym}
\usepackage{amsmath}
\usepackage{amssymb}
\usepackage{bm}
\usepackage{wasysym}
\usepackage{ulem}
\usepackage{braket}
\usepackage{hyperref}

\begin{document}
 \title{Entanglement in a driven two-qubit system coupled to common cavity}

\author{Amit Dey}
\affiliation{Department of Physics, Ramananda College, Bankura University, India 722122.}
\email{amit.dey.85@gmail.com }
\date{\today}
\begin{abstract}
A system, comprised of a qubit pair coupled to a common cavity, is studied with the aim of establishing qubit entanglement. This study is the follow up of the paper Phys. Rev. A {\bf 111}, 043705 (2025), where similar model was investigated for an initially vacuum cavity. In the present manuscript the cavity with finite initial occupancy is considered and the effect of asymmetric qubit-cavity couplings is investigated. For a closed system scenario, the ratio of the qubit-cavity couplings shows a threshold beyond which no maximally-entangled qubit state is available. The threshold value is shown to depend on the excitation level of the cavity. For a driven-dissipative case steady-state entanglement is shown to depend non-monotonically on the qubit drive. Intricate interplay of drive, dissipation, and coupling asymmetry is shown to be pivotal for steady-state entanglement generation.  
\end{abstract}

\maketitle

\section{Introduction}\label{intro}
Quantum entanglement is an indispensable ingredient for quantum information processing and has no counterpart in classical physics \cite{Nielsen2010-zt,PhysRev.47.777,PhysicsPhysiqueFizika.1.195, RevModPhys.81.865}. Entanglement finds its applications in multitude of sub-fields such as quantum key distribution \cite{RevModPhys.81.1301,PhysRevLett.67.661,nadlinger2022experimental}, quantum communication \cite{Gisin2007,chen2021integrated,yuan2010entangled,nauerth2013air}, quantum control\cite{li2023optimal,PhysRevA.75.012330,MORANDI2024129390,goodwin2022adaptive}, and quantum biology \cite{sarovar2010quantum,PhysRevLett.106.040503}.

The importance of entanglement has led to development of theoretical and experiment frameworks to achieve robust entangled qubit pair \cite{shankar2013autonomously,oh2006entanglement,egger2019entanglement,gonzalez2014generation,poletto2012entanglement,PhysRevLett.79.1,PhysRevLett.85.2392,jiang2024generating,chen2003generation,gomez2021entangling}. Photon-mediated entangled qubits are vital for quantum gate operation in cavity-QED platforms and has been utilized to design scalable architectures for quantum information processing \cite{noh2023strong,PhysRevA.77.023818,mckay2015high,srinivasa2024cavity,PhysRevB.107.155302,PhysRevApplied.6.064007,PhysRevB.104.115308}. In a large qubit system the cavity-qubit couplings might vary for various reasons. Distant qubits are engaged with spatially varying electromagnetic fields, giving rise to asymmetric cavity-qubit coupling \cite{PhysRevA.96.042714}. Parameter fluctuation of various circuit components results in fluctuating coupling strengths \cite{RevModPhys.93.025005}. Non-uniform couplings in a dual-species qubit architecture provide remarkable shield against the complicacies of qubit-specific read out \cite{PhysRevX.12.011040,PhysRevLett.128.083202,anand2024dual}. The availability of maximally entangled state (MES) has significant dependence on the asymmetry of couplings for various qubits. Multiple factors such as drive, dissipation, coupling asymmetry stage an intricate physics that decides the degree of steady-state entanglement \cite{dey2025correlating}. The photon and qubit degrees of freedom influence each other in cavity-QED systems providing avenues for state manipulation. The cavity state becomes determining factor for MES generation in related systems. Cavity-mediated qubit entanglement has been realized by using Fock states \cite{tessier2003entanglement,jin2006entanglement,cheng2022applications}, thermal states \cite{kim2002entanglement}, and coherent states \cite{jian2004entanglement,jarvis2009dynamics}. Such entanglement harvesting is not only limited to the usage of non-classical light. With appropriate pulse shaping, a classically driven cavity is very much able to generate entanglement between non-interacting qubits \cite{ahn2023qubit}. The qubit state also plays a role as a catalytic agent to improvise cavity state \cite{PhysRevResearch.6.023127}.

In the present manuscript, we investigate the role of coupling asymmetry in generating inter-qubit entanglement, while we consider various excitation levels of the cavity. The study is mainly a sequel of Ref. \cite{dey2025correlating}, where entanglement generation is investigated for various coupling ratios, when the cavity vacuum indulges qubit-qubit interaction via virtual photons. In a driven-dissipative set up an optimum range of qubit drive has also been sought \cite{dey2025correlating}. Here, we present an analytical treatment of a similar model with the presence of real photon in the cavity. We find a threshold coupling ratio for obtaining MES and 
an interesting dependence of the threshold value on the photon number present in the cavity. A non-monotonic dependence of steady-state entanglement on the qubit drive strength is found out, and complex interplay of drive, dissipation and coupling asymmetry is discussed. For a fixed drive, counter-intuitive appearance of finite entanglement is obtained for highly asymmetric cavity-qubit couplings.

The manuscript is arranged as follows. In Sec. \ref{Hamiltonian} we introduce our model and deduce an effective Hamiltonian for a dispersive case. In Sec. \ref{closed} we present numerical results related to entanglement dynamics and locate the threshold of coupling ratio for various cavity excitations. The results are supported by analytical calculations. In Sec. \ref{open} we present the driven-dissipative case and present an interesting interplay of coupling asymmetry, drive, and dissipation. Finally, we conclude and discuss further scopes of our study in Sec. \ref{conclude}.

\section{model Hamiltonian}\label{Hamiltonian}
Here we consider a system where a qubit pair is coupled to a common cavity mode. The Hamiltonian corresponding to our model is given by
\begin{eqnarray}
H_{\rm model}&=&\omega a^{\dagger}a+\epsilon\sum_{i=1}^2 s_i^z+\sum_{i=1}^2 g_i(a^{\dagger}s_i^-+{\rm h. c.}),
\end{eqnarray}
where $a$ and $s_i^-$ are destruction operators for the photon and i-th qubit degrees of freedom, respectively. $\omega$, $\epsilon$, and $g_i$ are photon frequency, qubit frequency, and qubit-photon coupling strength for i-th qubit, respectively. Now, we employ a canonical transformation which transforms the Hamiltonian to $H=e^{-X}H_{\rm model}~e^{X}$. Here, $X=\sum_j (g_j/\delta)(a^{\dagger}s_j^--{\rm h. c.})$ and $\delta=\epsilon-\omega$ measures the qubit-photon resonance. Considering terms up to second order we can expand $H$ as
\begin{eqnarray}
H &\approx& H_{\rm model}+[H_{\rm model},X]+\frac{1}{2!}{\Big[}[H_{\rm model},X],X{\Big ]}, \nonumber \\
&=& \omega a^{\dagger}a+\epsilon\sum_i s_i^z+\sum_i\frac{\tilde{g}_i}{\delta}s_i^z+\sum_{i\neq j}\frac{\tilde{g}_i\tilde{g}_j}{2\tilde{\delta}} (s_i^+s_j^-+ {\rm h. c.}), \nonumber \\
\label{effective}
\end{eqnarray}
where, $\tilde{g}_j=g_j\sqrt{2N_{\rm ph}+1}$ and $\tilde{\delta}=\delta (2N_{\rm ph}+1)$.
For $N_{\rm ph}$ photons, if we initiate the system to the state $|0N_{\rm ph}1\rangle$ or $|1N_{\rm ph}0\rangle$ the system remains in the subspace of $N_{\rm ph}+1$ excitations.  Here the first and third tags within the ket state designate the excitation status of the first and second qubits, respectively. $1(0)$ represents excited (ground) state of a particular qubit. $N_{\rm ph}$ in the second position denotes the photon excitations in the cavity. In the strong dispersive scenario $(\delta \gg g_{1,2})$ there is no exchange of real quanta between the qubit and the cavity. The qubit system is eventually limited to the subspace comprised of the states $|10\rangle_q$ and $|01\rangle_q$, where the first and the second positions in the ket are reserved for the first and second qubits, respectively.  Under this condition, we can safely eliminate the first two terms in Eq. \ref{effective} and write
\begin{eqnarray}
H&=&\sum_i\frac{\tilde{g}_i}{\delta}s_i^z+\sum_{i\neq j}\frac{\tilde{g}_i\tilde{g}_j}{2\tilde{\delta}} (s_i^+s_j^-+ {\rm h. c.}).
\label{effective1}
\end{eqnarray}
We see in Eq. \ref{effective1} that the renormalized qubit energy clearly depends on the cavity photon number and this in turn influences the qubit dynamics. The qubit-qubit interaction is realized in second-order perturbation via exchange of virtual photons \cite{PhysRevLett.85.2392,dey2025correlating}. Exchange of real photon also generates qubit entanglement when a resonant condition is considered \cite{dey2025correlating, ahn2023qubit}. Next, we analyze the entanglement dynamics for various values of $N_{\rm ph}$ and measure the role of coupling asymmetry $g_2/g_1$. The symmetric-coupling case is the situation when $g_2/g_1=1.0$.
\begin{figure}[t]
    \centering
    \includegraphics[scale = 0.35,angle=90,angle=90,angle=90]{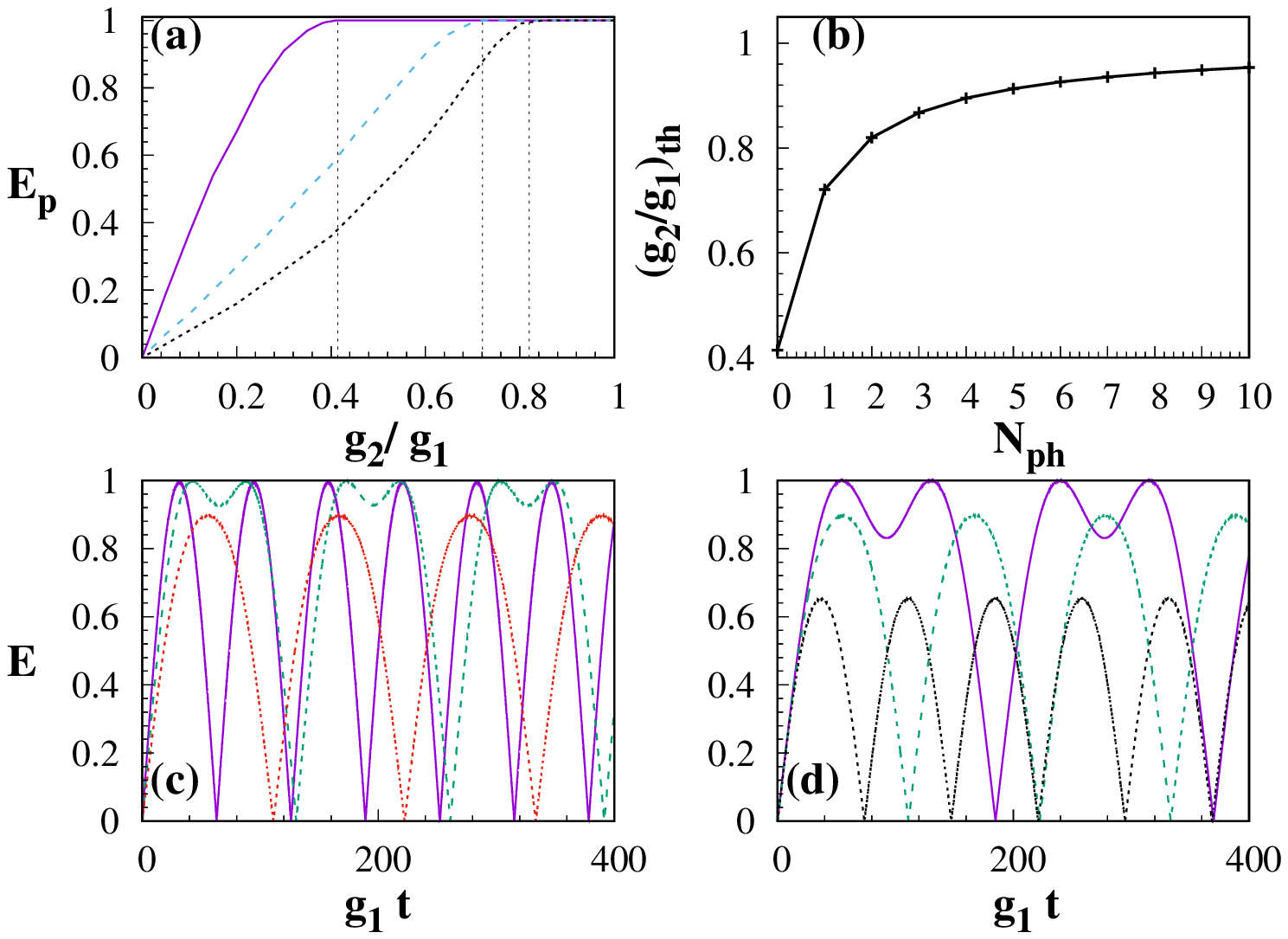}
    \caption{Entanglement is plotted against coupling asymmetry for various $N_{ph}$, when closed system is considered. (a) Peak entanglement is plotted against $g_2/g_1$ when $N_{ph}=0$ (solid), $1$ (dashed), $2$ (dotted). The vertical dotted lines from left to right mark the $(g_2/g_1)_{th}$ for $N_{ph}=0, 1, 2$, respectively. (b) $(g_2/g_1)_{th}$ is plotted against $N_{ph}$. (c) Entanglement dynamics for $g_2/g_1=1.0$ (solid), $0.8$ (dashed), and $0.6$ (dotted), when $N_{ph}=1.0$. (d) Entanglement dynamics for $N_{ph}=0$ (solid), $1.0$ (dashed), and $2.0$ (dotted), when $g_2/g_1=0.6$.}
    \label{fig1}
\end{figure}
\section{Closed system dynamics} \label{closed}
The effective two-qubit system is governed by the Hamiltonian Eq. \ref{effective1}. Throughout the paper we consider the initial state of the combined system as $|0N_{\rm ph}1\rangle$, i. e., $|01\rangle_q$ for the effective qubit system. The temporal behavior of the qubit system can be obtained by calculating ${\rm e}^{-iHt}|01\rangle_q$. The second term in the Hamiltonian Eq. \ref{effective1} transfers excitation between the qubits and eventually produces a superposition of $|10\rangle_q$ and $|01\rangle_q$ states. We note that the absolute values of the coefficients associated with $|01\rangle_q$ and $|10\rangle_q$ are equal for a MES. Exploiting this property we find that the necessary condition for MES is given by
\begin{eqnarray}
\frac{(\tilde{g}^2_2/\delta-\lambda_-)^2}{(\tilde{g}_1\tilde{g}_2/c\delta)^2}
+\frac{(\tilde{g}_1\tilde{g}_2/c\delta)^2}{(\tilde{g}^2_2/\delta-\lambda_-)^2}&\le& 6, \label{condition}
\end{eqnarray}
where $c=2N_{\rm ph}+1$ and the eigenvalues of Hamiltonian $H$ given by
\begin{eqnarray}
\lambda_\pm=\frac{\tilde{g}^2_1+\tilde{g}^2_2}{\delta}\pm\sqrt{\frac{(\tilde{g}^2_1+\tilde{g}^2_2)^2}{\delta^2}-4\frac{\tilde{g}^2_1\tilde{g}^2_2}{\delta^2}(1-1/c^2)}. \nonumber \\
\end{eqnarray}
The inequality in Eq. \ref{condition} produces the condition $g_2/g_1\ge [1/c+\sqrt{1+1/c^2}]^{-1}$ for obtaining MES. It is clear that the threshold value of $g_2/g_1$ for obtaining MES increases with greater cavity photon number $N_{\rm ph}$. It is evident from Fig. \ref{fig1} (a) and (b) that the threshold value $(g_2/g_1)_{\rm th}$ increases with increasing number of photons. The numerically obtained threshold values exactly match the theoretical values obtained from the expression $(g_2/g_1)_{\rm th}=1/c+\sqrt{1+1/c^2}]^{-1}$. Fig. \ref{fig1} (b) shows that at large $N_{\rm ph}$ values $(g_2/g_1)_{\rm th}$ approaches unity. This is also supported by the fact that, when $c>>1$ the expression of $(g_2/g_1)_{\rm th}$ approaches $1$. Fig. \ref{fig1} (c) shows that, for a fixed $N_{\rm ph}=1.0$ MES at peak values $E=E_p=1.0$ is available only when $g_2/g_1<(g_2/g_1)_{\rm th}$ (i. e., below threshold value $(g_2/g_1)_{\rm th}\approx 0.75$). Fig. \ref{fig1} (d) shows the availability of MES only for $N_{\rm ph}=0$ at fixed $g_2/g_1=0.6$. This is also reflected in Fig. \ref{fig1} (a) where $g_2/g_1=0.6$ falls within the threshold limit only for $N_{\rm ph}=0$. Therefore, the availability of MES prefers greater coupling symmetry when $N_{\rm ph}$ becomes larger. It is interesting to note that, for the resonant case $\delta=0$ and uniform coupling $g_1=g_2=g$, a vacuum cavity $N_{\rm ph}=0$ is unable to generate qubit entanglement, whereas, the entanglement peak varies as $1/N_{\rm ph}$ for excited cavity, when the system is initiated at $|0N_{\rm ph}0\rangle$ state \cite{ahn2023qubit}. This is markedly different from our present study, where we are able to generate qubit entanglement even with cavity vacuum and initial state $|001\rangle$ in a dispersive set up.  Moreover, the entanglement peak is always a MES for uniform coupling, irrespective of $N_{\rm ph}$. The entanglement in our study is essentially generated by exchange of virtual photons. Next, we deal with a driven-dissipative version of our model, where we employ coherent qubit drive to the second qubit only.
\begin{figure}[b]
    \centering
    \includegraphics[scale = 0.39,angle=90,angle=90,angle=90]{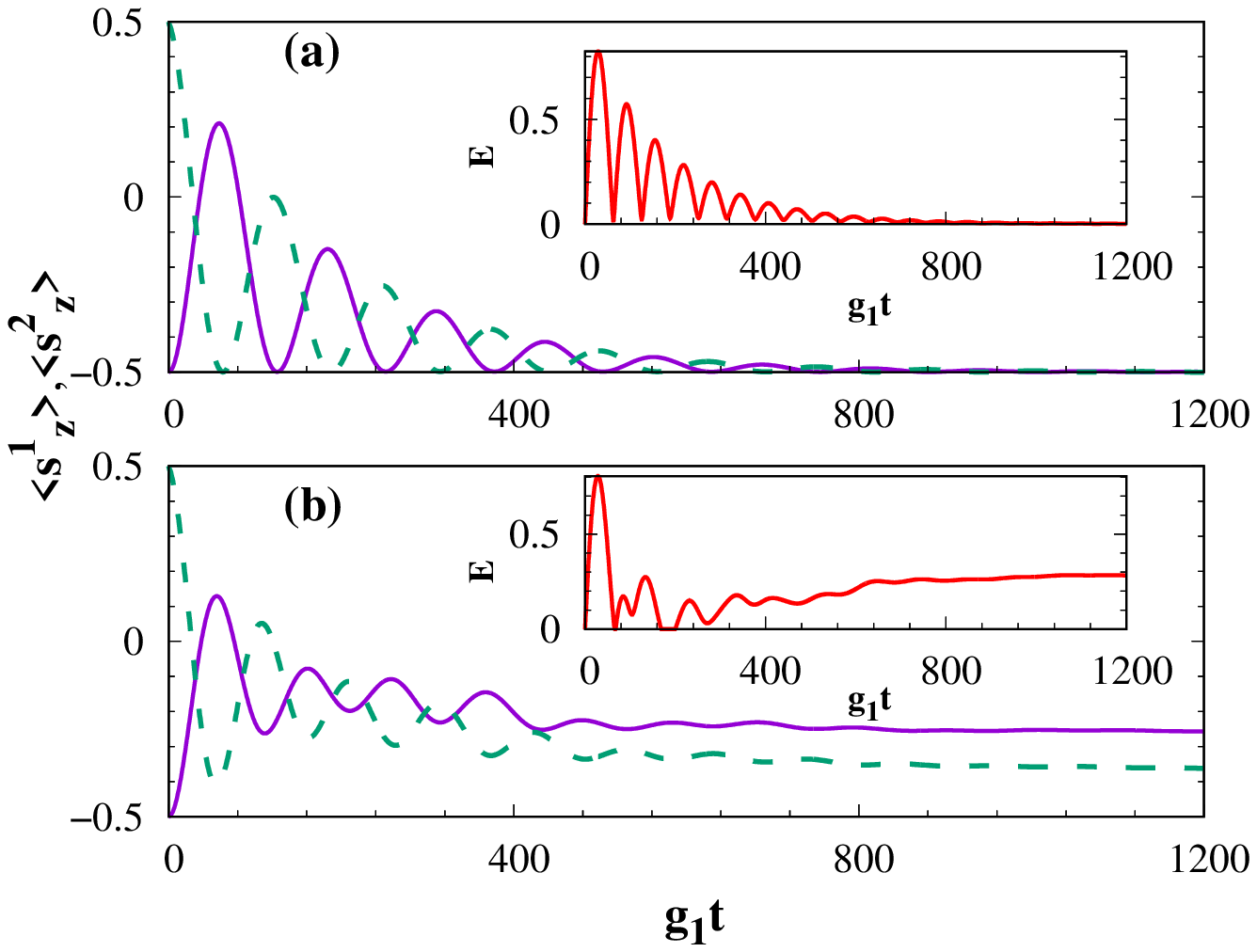}
    \caption{Qubit dynamics is plotted for $N_{\rm ph}=1.0$, $g_1=g_2$, and decay constants $\kappa=g_1$ and $\gamma=0.005g_1$ when (a) there is no qubit drive applied and (b) second qubit is driven with the drive strength $d=0.01g_1$. Average values of $s^1_z$ and $s^2_z$ are plotted by solid and dashed lines, respectively. The insets show entanglement dynamics attaining steady-state values. For case (a) $\omega=50g_1$ and $\epsilon=10g_1$ and for case (b) $\omega_d=9.99g_1$. $\omega$ and $\epsilon$ values are same as in (a). }
    \label{fig4}
\end{figure}
\section{Driven-dissipative dynamics}\label{open}
In a realistic cavity-QED system, there are various dissipative channels interfering with its unitary dynamics. Under the action of dissipation, the qubits eventually land in the ground state, and the cavity becomes vacuum. The open-system dynamics of the system is governed by the Lindblad master equation
\begin{eqnarray}
\dot{\rho}(t)&=&-i[H^\prime,\rho(t)]+\kappa\mathcal{L}[a]+\gamma\sum_j\mathcal{L}[s^-_j],
\end{eqnarray}
where, $\mathcal{L}[a]=1/2[2a^{\dagger}\rho(t)a-a^{\dagger}a\rho(t)-\rho(t)a^{\dagger}a$], $\kappa$ and $\gamma$ are dissipation rates for the cavity and qubit, respectively. The Hamiltonian in the rotating frame of qubit drive frequency is given by
\begin{eqnarray}
H^\prime &=& (\omega-\omega_d)a^{\dagger}a+(\epsilon-\omega_d)\sum_{i=1}^2 s^z_i
+\sum_{i=1}^2g_i(a^{\dagger}s^-_i+{\rm h. c.}) \nonumber \\
&&~~~~~~~~~~~~~~~+d(s^+_2+s^-_2),
\end{eqnarray}
where $\omega_d$ and $d$ are the drive frequency and the drive strength, respectively. Here, drive is applied to the second qubit only to minimize effort for parameter control. In Fig. \ref{fig4} (a), we see that the undriven case of $d=0$ leads to an unentangled steady state for the qubits. The qubits reach their respective ground states at long times. Therefore, dissipation destroys the entanglement generated by the cavity-qubit Hamiltonian. The driven case for $d=0.01g_1$ in Fig. \ref{fig4} (b) shows that the qubits attain superposition of ground and excited states at long times. The inset shows finite steady-state entanglement. The advantage of qubit drive is evident in this analysis. Now, we are curious about the range of drive strength that ensures finite steady-state entanglement. We are also interested in investigating the interplay of coupling asymmetry and drive strength while producing steady-state entanglement.
\begin{figure}[t]
    \centering
    \includegraphics[scale = 0.39,angle=90,angle=90,angle=90]{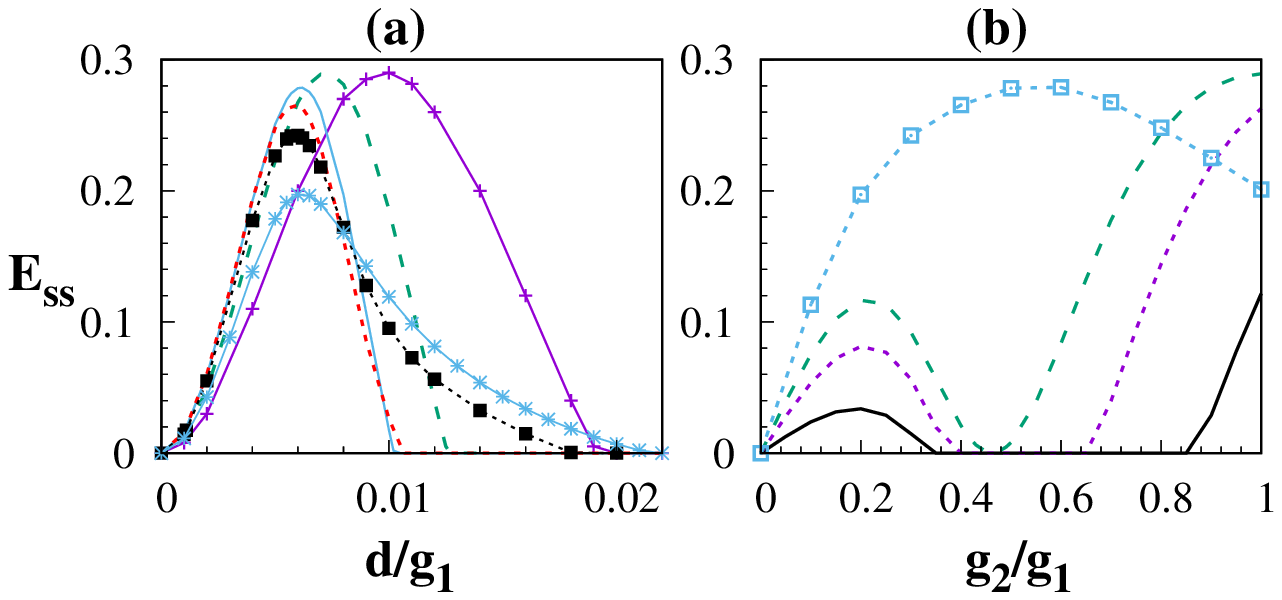}
    \caption{Steady-state entanglement is plotted for $N_{ph}=1.0$, when open system is considered with $\kappa=g_1$ and $\gamma=.005g_1$. (a) $E_{ss}$ is plotted against drive strength when $g_2/g_1=1.0$ (+), $0.7$ (dashed), $0.5$ (solid), $0.4$ (dotted), $0.3$ (filled square), and $0.2$ (*). (b) $E_{ss}$ is plotted against $g_2/g_1$ for $d/g_1=.016$ (solid), $0.012$ (dotted), $0.01012$ (dashed), and $.006$ (open square).}
    \label{fig2}
\end{figure} 

In Fig. \ref{fig2} we plot the steady-state entanglement for the driven-dissipative case. Fig. \ref{fig2} (a) shows that the steady state of the qubit pair remains entangled for an optimum range of qubit drive strengths and this holds true for various couplings $g_2/g_1$. Initially the steady-state entanglement $E_{ss}$ increases with the qubit drive and decreases after reaching a peak. Although the drive augments the inter-qubit correlation by counteracting qubit dissipation (as can be seen from Fig. \ref{fig4}), the inter-qubit coupling (which depends on $g_1g_2$ according to Eq. \ref{effective1}) is unable to adapt driving time scale beyond a limit of drive strength. 
\begin{figure}[t]
    \centering
    \includegraphics[scale = 0.35,angle=90,angle=90,angle=90]{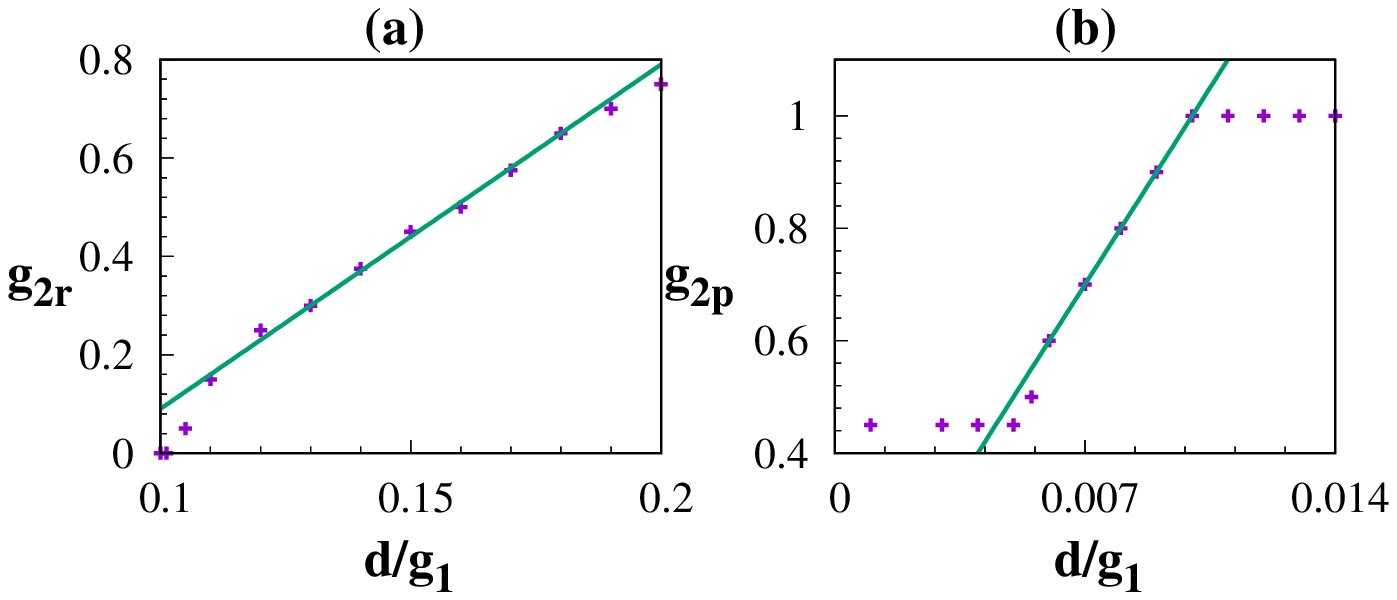}
    \caption{(a) The range of $g_2$ resulting $E_{ss}=0$ valley (as discussed in Fig. \ref{fig2} (b)) is plotted against $d/g_1$. Here $g_{2r}$ is the difference between maximum and minimum $g_2/g_1$ values producing $E_{ss}=0$. (b) The $g_{2p}=g_{2}/d_1$ values corresponding to the peak $E_{ss}$ (as shown in Fig. \ref{fig2} (b)) is plotted against $d/g_1$. Here $N_{ph}=1.0$, $\kappa=g_1$ and $\gamma=.005g_1$. The solid straight lines in both the figures present straight-line fitting for the data obtained in the specific regions.}
    \label{fig5}
\end{figure}
This is evident from Fig. \ref{fig2} (a) where the peak position shifts towards lower $d$ values for decreasing $g_2/g_1$. This observation can also be made for a case when the system is only comprised of coupled qubits with no photonic degrees of freedom.  The drive strength, where the peak appears, finally saturates around $d/g_1=0.06$. The peak value of $E_{ss}$ decreases with decreasing $g_2/g_1$ (i. e., with elevated coupling asymmetry). This means that the most favorable qubit drive (in terms of higher $E_{ss}$) prefers greater coupling symmetry. We also observe that the range of drive $d_r$ producing non-zero $E_{ss}$ apparently shrinks with decreasing $g_2/g_1$. Interestingly, at some lower $g_2/g_1$ values ($g_2/g_1=0.4$ in Fig. \ref{fig2} (a)) $d_r$ starts increasing with decreasing $g_2/g_1$. This feature is exclusive for a photon-coupled-qubit system and absent in directly coupled qubit system 
. This feature directly results in a hump of $E_{ss}$ at lower $g_2/g_1$ values in Fig. \ref{fig2} (b) for a particular $d/g_1$. This non-monotonic behavior is kind of entanglement reappearance in coupling domain. Fig. \ref{fig2} (b) presents a number of interesting observations. First, for a fixed $d$ there is a valley in steady-state entanglement $E_{ss}$ followed by a hump at lower $g_2/g_1$ values. Secondly, we also observe that $E_{ss}=0$ region decreases with decreasing $d$ and at some point it vanishes becoming a dip (as the $d/g_1=.01012$ case in Fig. \ref{fig2} (b)). This fact can also be connected to the variation of $g_{2r}$ with $d/g_1$ in Fig. \ref{fig5} (a), which indicates linear variation of the $E_{ss}=0$ valley width (in $g_2/g_1$ domain) with varying qubit drive. The dip in Fig. \ref{fig2} (b) becomes shallower with reducing $d$ and the plot turns out to be parabola shaped (such as $d=.006g_1$ case in Fig. \ref{fig2} (b)) at lower $d$ values. It is interesting to observe that the maximum $E_{ss}$ for $d=.006g_1$ in Fig. \ref{fig2} (b) does not correspond to uniform coupling $g_2/g_1=1$. For low values of qubit drive, stronger effective interaction between the qubits does not pick up the drive efficiently. This requires asymmetric coupling $g_2/g_1<1$ (such as $g_2=.6g_1$ for $d=.006g_1$) so that the inter-qubit interaction becomes compatible with the driving time scale. 

\begin{figure}[b]
    \centering
    \includegraphics[scale = 0.39,angle=90,angle=90,angle=90]{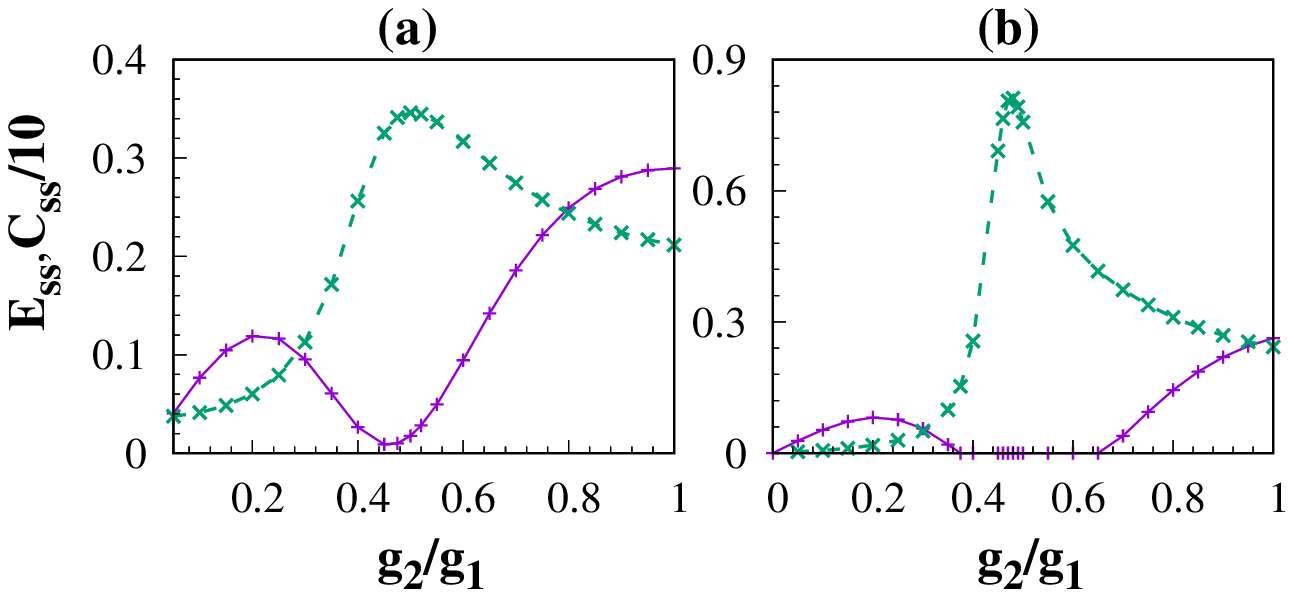}
    \caption{Steady-state values of entanglement ('+' markers) and cross correlation $C_{ss}/10$ ('x' markers) are plotted against coupling asymmetry for $N_{ph}=1.0$, when open system is considered with $\kappa=g_1$ and $\gamma=.005g_1$. (a) $d=0.01g_1$. (b) $d=.012g_1$.}
    \label{fig3}
\end{figure}

This explanation can also be related to the observation in Fig. \ref{fig5}(b). Fig. \ref{fig5} (b) shows that the $g_2/g_1$ corresponding to the peak value of $E_{ss}$ saturates to $g_2/g_1=1.0$ for larger drive $d/g_1$, whereas, it saturates to $g_2/g_1=0.45$ for lower drive. In the intermediate range of $d/g_1$ values $g_{2p}$ linearly varies with the drive in Fig. \ref{fig5} (b) indicating proportionality between drive strength and the most favorable coupling asymmetry.   The appearance of the hump in Fig. \ref{fig2} (b) directly follows from the enhanced $d_r$ at lower $g_2/g_1$ values in Fig. \ref{fig2} (a). To explain this, we take resort to the cross-correlation function, which is  
consodered as a witness of quantum correlation \cite{PhysRevA.71.032304, PhysRevA.96.052311, rice2006steady}.
 In Fig. \ref{fig3} we plot the steady-state entanglement along with the average cross correlation function $C_{ss}$ against the coupling asymmetry. Here the cross-correlation function is defined as 
\begin{eqnarray}
C^i_{ss}&=&\frac{\langle s^z_ia^{\dagger}a \rangle}{\langle s^z_i\rangle\langle a^{\dagger}a\rangle}, \\
C_{ss}&=&\frac{1}{2}\sum_{i=1}^2 C^i_{ss}.
\label{cross_corr}
\end{eqnarray}
We see in Fig. \ref{fig3} that $C_{ss}$ gets enhanced and acquires a peak at intermediate $g_2/g_1$ values, where $E_{ss}$ has a dip. More the qubits entangle with the photonic degrees of freedom lesser becomes the inter-qubit entanglement. When we deal with uniform coupling simultaneous excitation of photonic and qubit degrees of freedom is less probable, whereas, qubit-qubit correlation is more likely (due to effective qubit interaction $\sim g_1g_2$). As the coupling becomes non-uniform, the likelihood of {simultaneous excitation of photonic and qubit degrees of freedom is facilitated till the peaks appear in Fig. \ref{fig3}}. In this regime of $g_2/g_1$, $E_{ss}$ keeps on decreasing and finally bottoms to the valley of no entanglement. Now, reducing $g_2/g_1$ should also cost the local interaction between the driven qubit and photon, which results decreasing $C_{ss}$ below certain limit of $g_2/g_1$ value. This in turn builds up qubit-qubit correlation resulting the hump in the lower $g_2/g_1$ region. Finally, when $g_2/g_1$ becomes very low it is very hard to sustain any correlation in the steady state. Naturally, $E_{ss}$ vanishes as $g_2/g_1=0$ is approached.  
\section{Conclusion and discussion}\label{conclude}
In this paper we have studied the impact of cavity excitation on the entanglement generation in a qubit pair.  We find out a threshold value of coupling asymmetry, which dictates the availabily of MES depending on the cavity excitation level. An intricate interplay among qubit drive, dissipation, and coupling asymmetry is presented in the context of extracting steady-state entanglement. Given the minimal arrangement of various components and well-developed cavity-QED platform, the model can be efficiently implemented. The study is important to assess the robustness of generated qubit entanglement when cavity-qubit coupling varies in realistic situations. Moreover, it helps in optimizing qubit drive that stabilizes steady-state entanglement. 

The study motivates investigation of multipartite entanglement in a similar set up\cite{PhysRevLett.87.230404, PhysRevLett.90.027903}. We have investigated a dispersive case where virtual photon exchange builds up qubit correlation. A resonant case with driven cavity is another interesting situation when entanglement is contributed by real photon exchange processes \cite{ahn2023qubit, PhysRevA.96.052311}.
\bibliography{bibliography.bib}
\end{document}